\documentclass[twocolumn,prb,showpacs,amsmath,amssymb,longbibliography]{revtex4-1}

\usepackage{graphicx}
\usepackage{dcolumn}
\usepackage{bm}
\usepackage{tabularx}
\usepackage{multirow}

\begin{document}

\title {Prediction of the Weyl semimetal in the orthorhombic MoTe$_2$}

\author{Yan Sun$^1$}
\author{Shu-Chun Wu$^1$}
\author{Mazhar N. Ali$^2$}
\author{Claudia Felser$^1$}
\author{Binghai Yan$^{1,3,4,5}$}
\email{yan@cpfs.mpg.de}
\affiliation{$^1$Max Planck Institute for Chemical Physics of Solids, 01187 Dresden, Germany}
\affiliation{$^2$IBM Almaden Research Center, San Jose, California 95120, USA}
\affiliation{$^3$Max Planck Institute for the Physics of Complex Systems, 01187 Dresden,Germany}
\affiliation{$^4$ School of Physical Science and Technology, ShanghaiTech University, Shanghai 200031, China}
\affiliation{$^5$CAS-Shanghai Science Research Center, Shanghai 201203, China}

\date{\today}

\begin{abstract}
We investigate the orthorhombic phase (Td) of layered transition-metal-dichalcogenide MoTe$_2$ as the Weyl semimetal candidate. 
MoTe$_2$ exhibits four pairs of Weyl points lying slightly above ($\sim$ 6 meV) the Fermi energy in the bulk band structure.
Different from its cousin WTe$_2$ that was predicted to be a type-II Weyl semimetal recently, the spacing between each pair of Weyl 
points is found to be as large as 4 percent of the reciprocal lattice in MoTe$_2$ (six times larger than that of WTe$_2$). 
When projected to the surface, Weyl points are connected by Fermi arcs, which can be easily accessed by ARPES due to the large Weyl point separation. In addition, we show that the correlation effect or strain can drive MoTe$_2$ from type-II to type-I Weyl semimetal. 

\end{abstract}


\maketitle

\section{introduction}
The Weyl semimetal (WSM) is a topological semimetal~\cite{volovik2007quantum, Wan2011, Burkov:2011de}, in which bands disperse linearly in three-dimensional 
(3D) momentum space through a node, called Weyl point (WP). The WP acts as a monopole with fixed chirality~\cite{Nielsen1981}, 
a source or a sink of the Berry curvature. Similar to those of a topological insulator (TI), topologically protected surface states exists on the surface of a WSM.
Topologically different from ordinary Fermi surfaces (FSs), these surface states present unclosed FSs, called Fermi arcs, which connect 
the surface projections of WPs with the opposite chirality~\cite{Wan2011}. 
WSMs also exhibit exotic quantum transport phenomena, such as the chiral 
anomaly~\cite{Adler1969,Bell1969, bertlmann2000anomalies} characterized by the negative longitudinal magnetoresistance (MR)~\cite{Nielsen1983},
anomalous Hall effect ~\cite{Xu2011,Yang2011QHE,Burkov:2011de,Grushin2012} and nonlocal transport properties~\cite{Parameswaran2014,Zhang:2015ub}.

Recently the first WSM materials (Ta,Nb)(As,P) have been discovered by addressing the Fermi arcs in angle-resolved photoemission 
spectra (ARPES)~\cite{Xu2015TaAs,Lv2015TaAs,Yang:2015TaAs,Lv2015TaAsbulk,Liu2015NbPTaP,Xu2015NbAs,Xu:2015TaP}, 
which was originally predicted by band structure calculations~\cite{Weng:2014ue,Huang:2015uu}. 
Meanwhile a great amount of efforts  have also been devoted 
to their magneto-transport properties~\cite{Shekhar2015,Shekhar2015TaP,Huang2015anomaly,Zhang2015ABJ,Wang:2015wm,Yang:2015vz,Du:2015TaP},
such as extremely large, positive transverse MR~\cite{Shekhar2015} and large, negative longitudinal MR~\cite{Shekhar2015TaP,Du:2015TaP}. 
These family of WSMs exhibit ideal Weyl cones in the bulk band structure, i.e. the FS shrinks into a point at the WP.
Very recently, Soluyanov \textit{et. al.} proposed a new type of WSM in the compound WTe$_2$~\cite{Ali:2014bx}, referred as type-II WSM~\cite{Soluyanov:2015WSM2}.
Herein, the Weyl cone is strongly tilted so that WP exists at the touching point of the electron and hole pockets in the FS,
which may indicate salient response to the magnetic field. Topological Ferim arcs was demonstrated between a pair of WPs that 
are separated by $\sim 0.7 \%$ of the reciprocal lattice vector length and lie about 50 meV above the Fermi energy ($E_F$).

\begin{figure}[htbp]
\begin{center}
\includegraphics[width=0.45\textwidth]{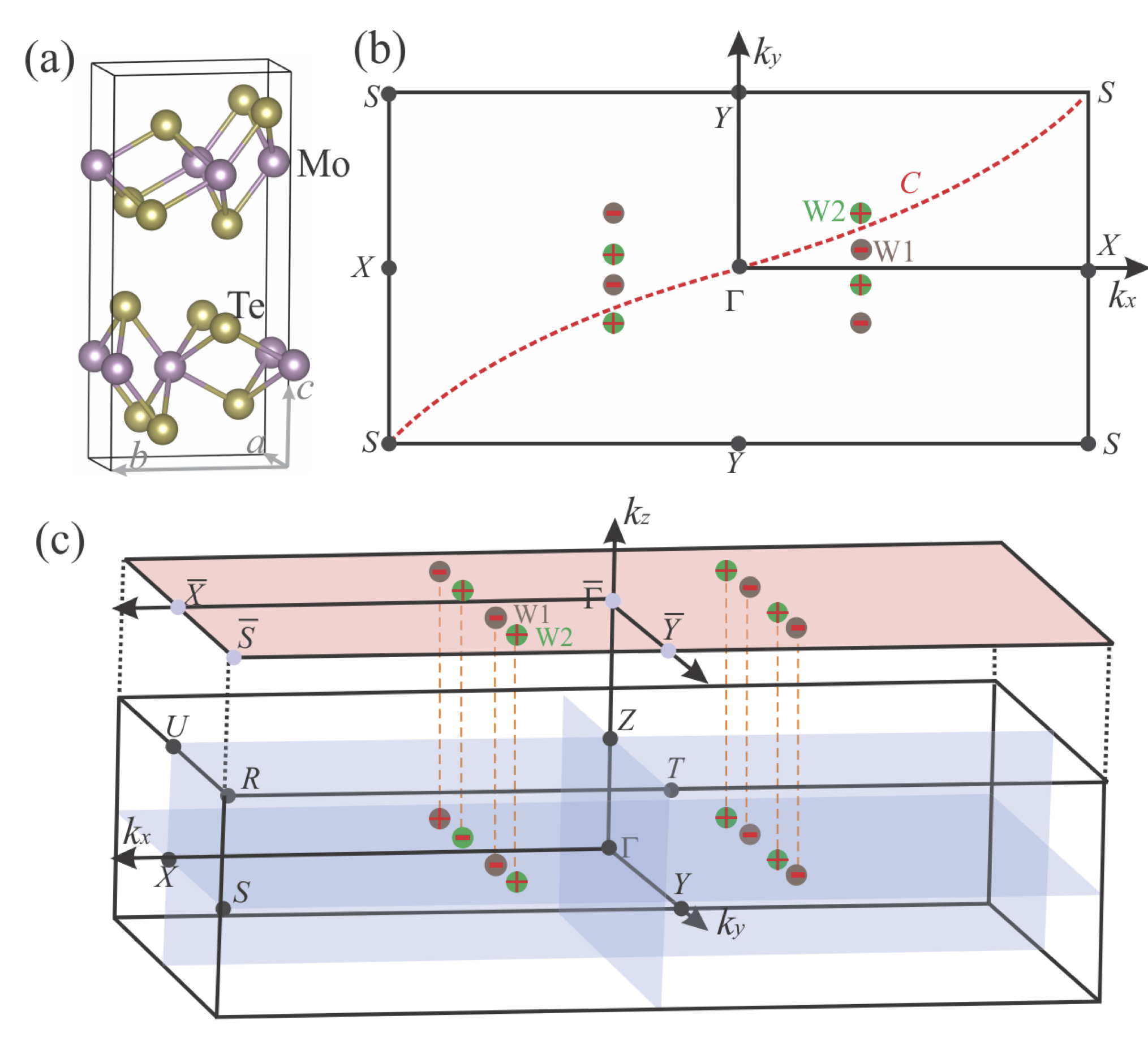}
\end{center}
\caption{
(color online)  (a) Orthorhombic crystal lattice structure of Td-MoTe$_{2}$ in the space group of Pnm2$_{1}$. 
(b) Brillouin zone (BZ) in the $k_z=0$ plane. WPs with positive and negative chiralities are
marked as green and gray dots. The evolution of Wannier charge centers between $\Gamma$ to $S$ point is calculated  along the red
curve $C$. (c) 3D bulk BZ and the projected surface BZ to (001) plane.
}
\label{lattic}
\end{figure}

\begin{figure*}[htbp]
\begin{center}
\includegraphics[width=0.85\textwidth]{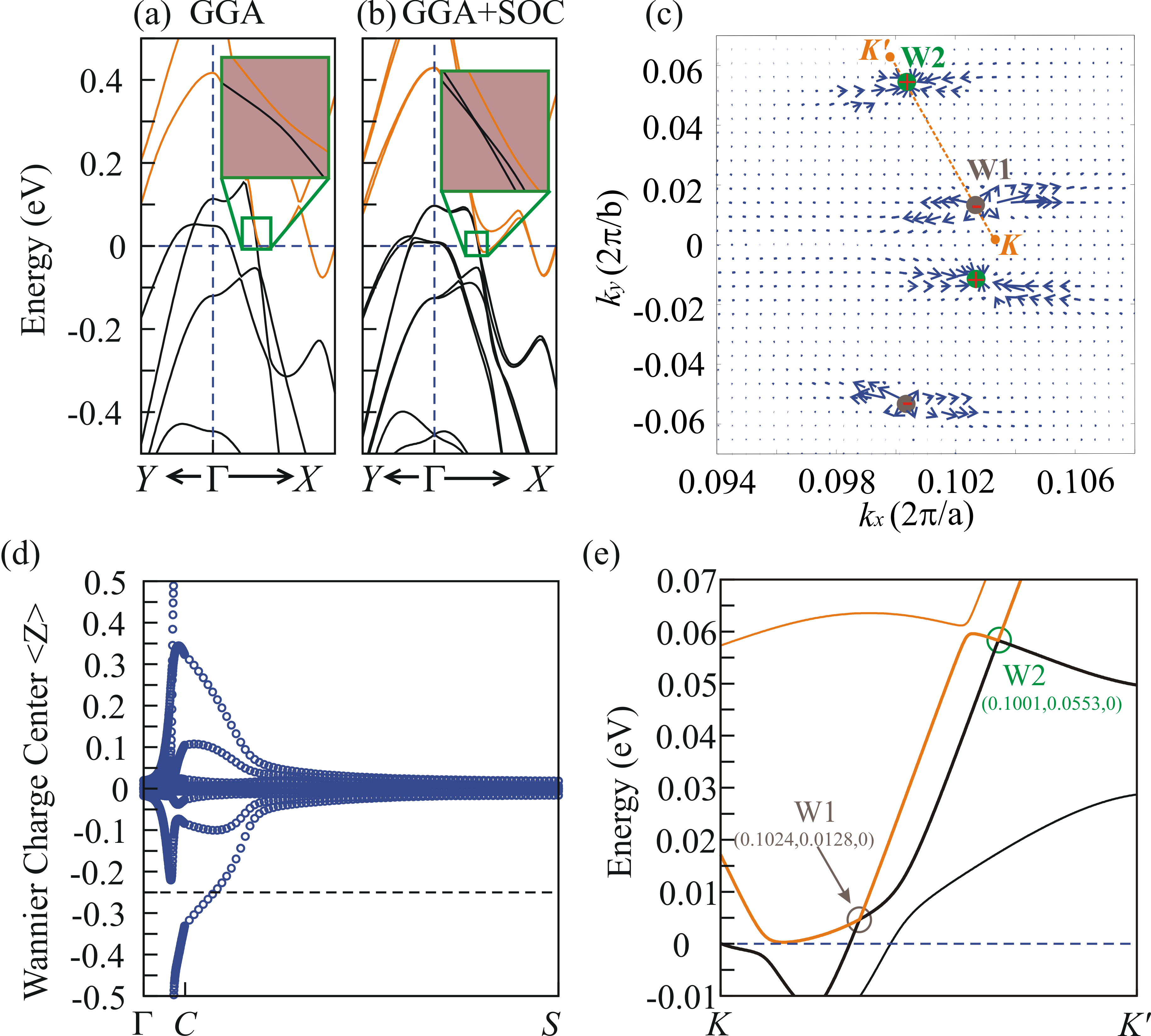}
\end{center}
\caption{
Bulk band structure around $\Gamma$ point in the direction of $Y-{\Gamma}-X$ direction (a) without and
(b) with the inclusion of SOC. (c) Berry curvature in the $k_z=0$ plane around two types of WPs.
The size  and directions of arrows represent the magnitude and orientations of the Berry curvature, respectively. 
WPs with positive and negative chirality are denoted by the green and gray points. (c) The evolution of Wannier
charge centers between two time-reversal-invariant points of $\Gamma$ and $S$ along the $C$ curve (see Fig. 1b).
The Wannier charge centers cross the reference horizontal line once, indicating the nontrivial $Z_2$ invariant in the 
$C$-$k_z$ plane. (e) Band structure crossing two types of WPs. The two points of $K$ and $K'$ is shown in (c).
Black and orange bands in (a), (b) and (e) are the lowest N-th bands and other higher bands bands, respectively.
}
\label{bulk}
\end{figure*}

\begin{figure*}[htbp]
\begin{center}
\includegraphics[width=0.85\textwidth]{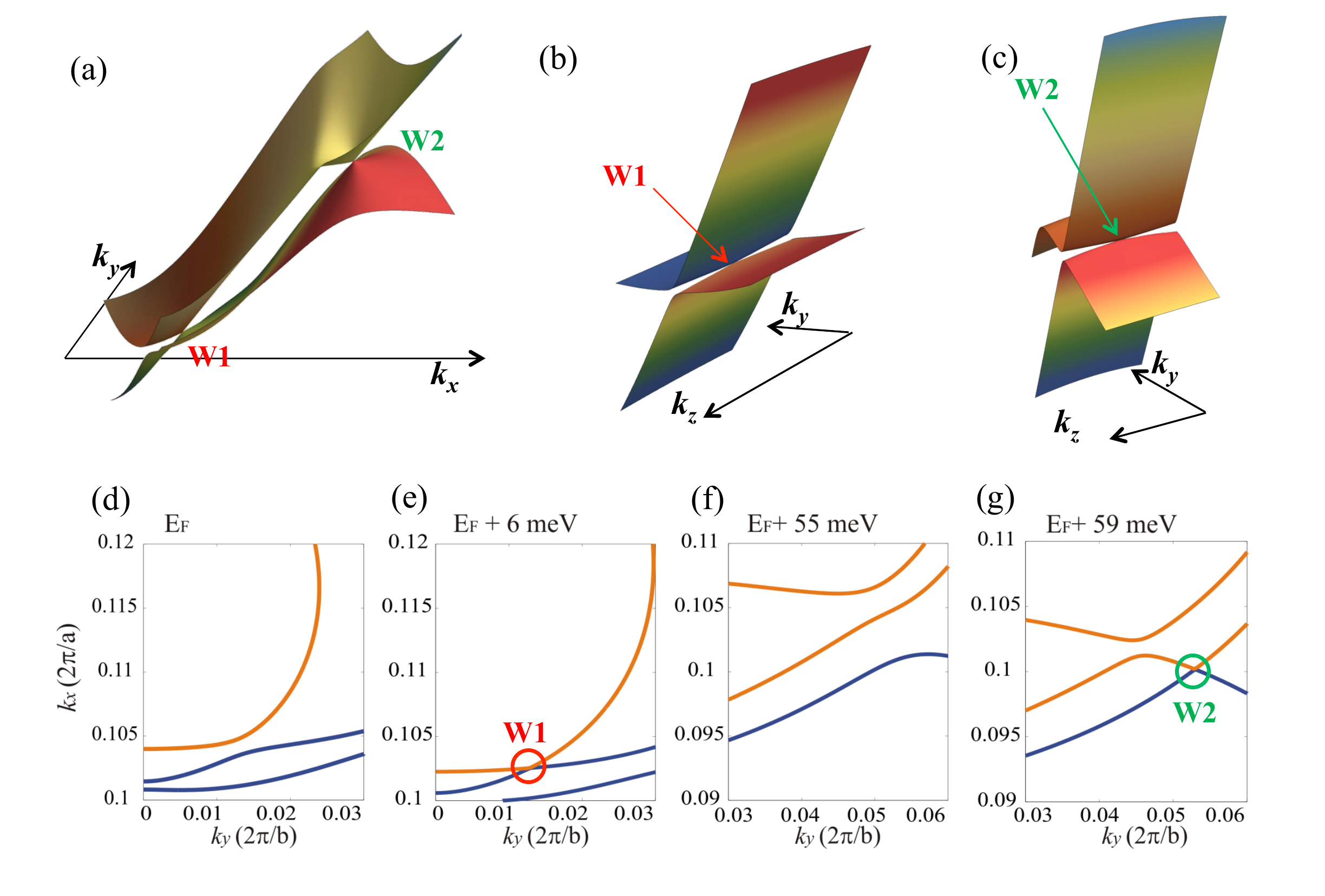}
\end{center}
\caption{
3D plot of WPs in (a) $k_z=0$, (b) $k_x=0.1024\frac{2\pi}{a}$ and (c) $k_x=0.1001\frac{2\pi}{a}$ planes. 
(d-g) The evolution of FS around two WPs in $k_z=0$ plane. WPs related
Fermi surface linear crossing can be at $E_F$ + 6 meV and $E_F$ + 59 meV, respectively.
The orange and blue FSs come from the electron and hole pockets.
}
\label{3D-WP}
\end{figure*}

\begin{figure*}[htbp]
\begin{center}
\includegraphics[width=0.85\textwidth]{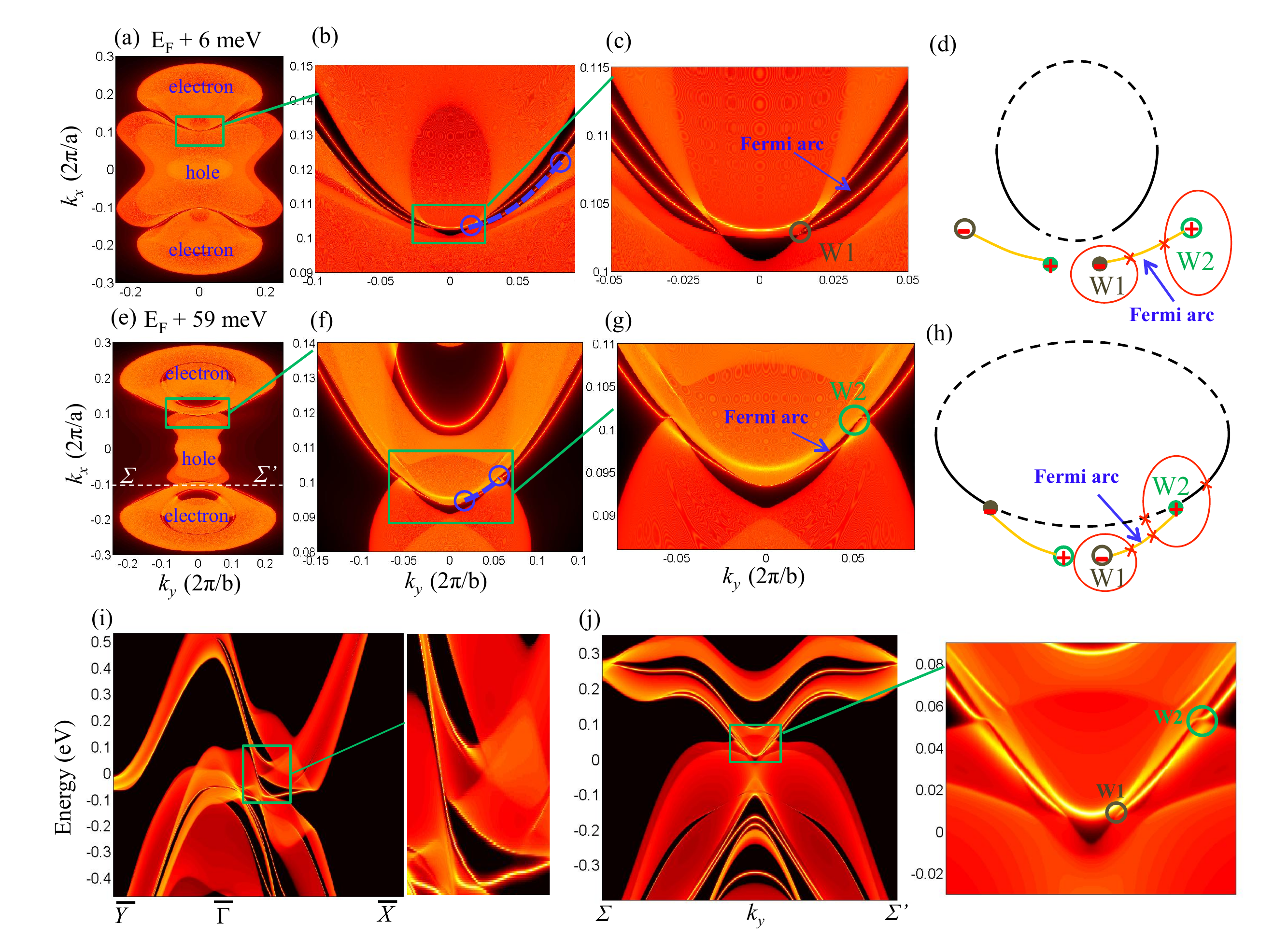}
\end{center}
\caption{Fermi surfaces and surface band structures. (a-c) Surface FS and (d) the illustration of Fermi arcs at $E_F$ + 6 meV.
(f-g) Surface FSs and (h) corresponding schematics at $E_F$ + 59 meV.
(i) Surface energy dispersion along the high symmetry lines of $\overline Y$-$\overline{\Gamma}$-$\overline X$.
(j) Surface energy dispersion along the $\Sigma - \Sigma'$ crossing WPs.
Brighter colors represent the higher LDOS. The exact projections of W1 and W2 are denoted as filled grey and green dots.
The end points of Fermi arcs are marked as empty circles. The Fermi arcs are highlighted by dashed lines in (b) and (f). 
}
\label{surface}
\end{figure*}

\begin{figure*}[htbp]
\begin{center}
\includegraphics[width=0.85\textwidth]{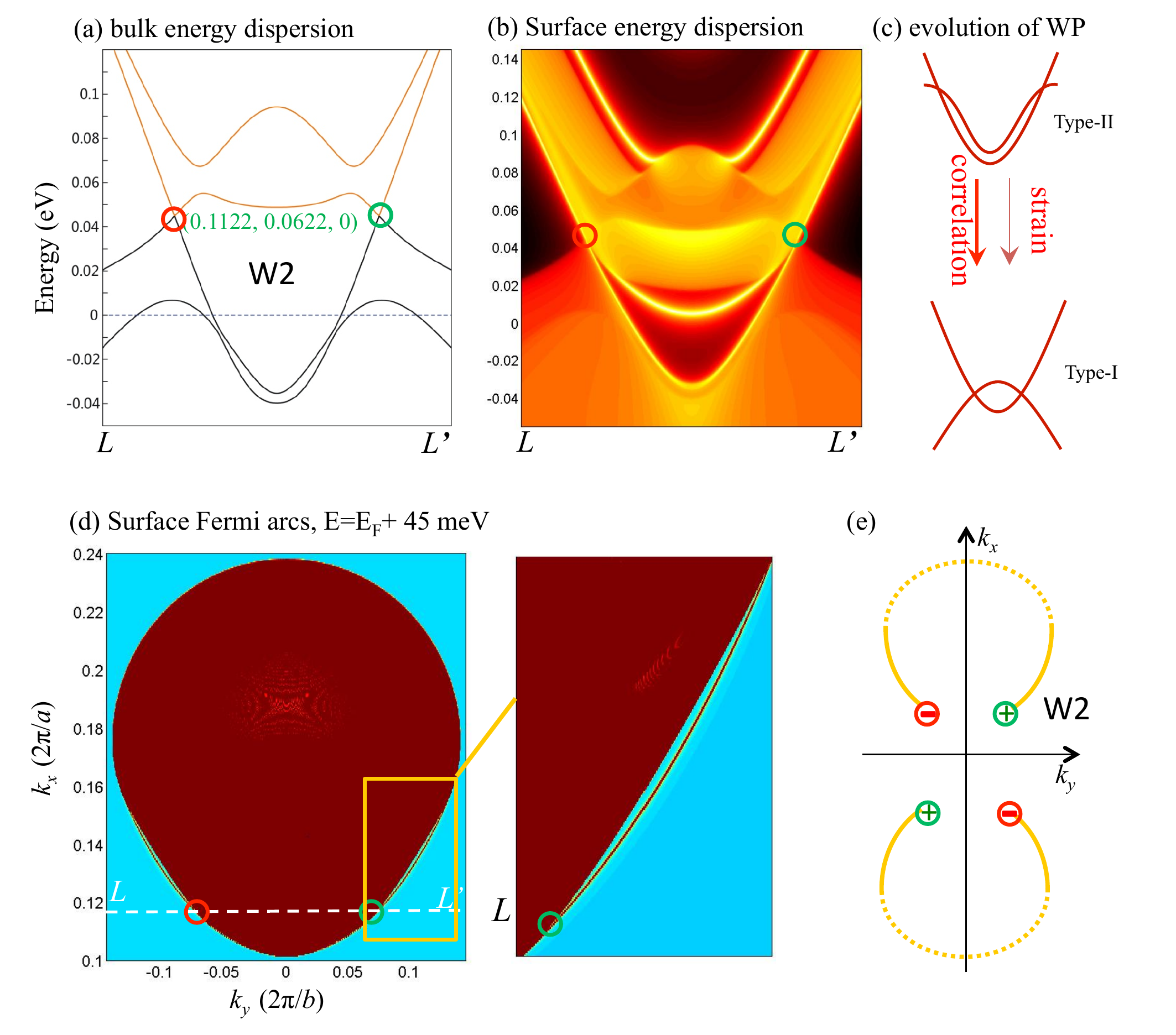}
\end{center}
\caption{Electronic structure calculated from hybrid functional. (a) Bulk band structure along two WPs in $k_y$  directions,
and (b) corresponding surface energy dispersion. (c) Schematic of WP evolution with correlation and compression strain.
(d) Surface FS at the energy level of $E=E_F+45$ meV and (e) corresponding 
schematics for the surface Fermi arcs. The two points of $L$ and $L'$ are shown in (d). 
The red and green circles represent WPs with opposite chirality.
}
\label{hse}
\end{figure*}

Provoked by WTe$_2$~\cite{Soluyanov:2015WSM2}, we predict that the orthorhombic phase of MoTe$_2$, which was synthesized very recently~\cite{Qi2015MoTe2}, 
is a new type-II WSM candidate. The layered transition metal dichalcogenide MoTe$_2$ can crystalize into three phases in 
different experimental conditions: the 2H, 1T$^\prime$ and Td phases. The 2H phase is semiconducting, in which the Mo 
atom has trigonal prismatic coordination with the Te atoms. The 1T$^\prime$ (also called $\beta$-phase)~\cite{Brown1966} and 
Td phases are semimetallic and exhibit pseudo-hexagonal layers with zig-zag metal chains. The 1T$^\prime$ phase is a monoclinic 
lattice that is stable at room temperature, arising from the slight sliding of layer-stacking of the orthorhombic lattice of the Td phase (see Fig. 1a).
The Td compound can be obtained by cooling the 1T$^\prime$ phase down to 240 K~\cite{Qi2015MoTe2,Hughes1978,Zandt2007}.
We note that the 1T$^\prime$ structure exhibits inversion symmetry (space group $P12_1/m1$, No. 11) while the Td one does not (space group $Pmn2_1$, No. 31).
The 1T$^\prime$ phase of MoTe$_2$ was speculated to be a WSM candidate as a pressurized WTe$_2$ in Ref.~\onlinecite{Soluyanov:2015WSM2}.
We note that, however, the 1T$^\prime$ phase cannot be a WSM simply because of the co-existence of time-reversal symmetry (TRS) and the inversion symmetry,
because the WSM requires the breaking of at least  one of the TRS and inversion symmetry.
Therefore, only the Td phase, which is isostructural to WTe$_2$, can be a possible WSM.

In this work, we have investigated the Td-MoTe$_2$ as a WSM candidate by $ab~initio$ density-functional theory (DFT) calculations.
We indeed find four pairs of WPs in the band structure (see Fig. 1c), similar to WTe$_2$.
Each pair of WPs exhibit considerable spacing in the Brillouin zone, $\sim 4.2 \%$ of the reciprocal lattice vector,
in which one WP exists merely 6 meV above $E_F$ and the other WP lies 59 meV above $E_F$.
We observe clear Fermi arcs in the surface state calculations.
Due to the large momentum spacing and the close vicinity to $E_F$ of WPs,
we expect that the Fermi arcs and other interesting quantum phenomena of MoTe$_2$ will be easily accessed by future ARPES and transport experiments.

\section{Methods and crystal structure}
DFT calculations were performed with the Vienna $Ab~initio$ Simulation Package (\textsc{vasp})~\cite{kresse1996}
with projected augmented wave (PAW) potential. The exchange and correlation energy was considered in both
generalized gradient approximation (GGA)~\cite{perdew1996} level with Perdew-Burke-Ernzerhof (PBE) functional and
hybrid functional (HSE06) ~\cite{HSE06}. 
The energy cutoff was set to 300 eV for a plane wave basis.  The tight-binding matrix was constructed by projecting
the Bloch states into maximally localized Wannier functions (MLWFs)~\cite{Mostofi2008}. 
We employed the experimental lattice parameters and atomic positions measured by our recent experiment~\cite{Qi2015MoTe2}. 
For completeness, the lattice constants are $a=3.477 \AA$,  $b=6.335 \AA$, and $c=13.883 \AA$.
There are two Mo atoms and eight Te atoms in the primitive unit-cell.
All the atoms locate at the $2a$ sites of the $Pmn2_1$ space group  with reduced positions of Mo: (0.0, 0.60520, 0.50034), (0.0, 0.03010, 0.01474), and (0.0, 0.86257, 0.65574), (0.0, 0.64045, 0.11246), (0.0, 0.28997, 0.85934), (0.0, 0.21601, 0.40272).

\section{Results and discussions}
The Td-MoTe$_2$ exhibits a semimetallic feature in the band structure, as shown in
Figs. 2a and 2b. 
We refer to the lowest $N$-th bands (black bands) as the valence bands of a 
semimetal and the other higher bands (orange bands) as the conduction bands at each $k$-point, 
where $N$ is the number of total valence electrons.
When SOC is taken into consideration, the spin degeneracy is lifted up in the band structure due to the lack of inversion symmetry. 
A tiny direct gap between the conduction and valence bands exist at the one fifth position of the ${\Gamma}-X$ line. However, near this point a pair of linearly band-crossing points exists aside from the high symmetry ${\Gamma}-X$ line, 
as shown in Fig. 2e. One point is merely 6 meV above $E_F$ (labeled as W1) and the other is 59 meV above $E_F$ (labeled as W2).

To validate the topology of W1 and W2 band-crossing points,
we investigate the Berry curvature in the $k_z = 0$ plane calculated from the tight-binding Hamiltonian with MLWFs.
The Berry curvature is integrated over all the $N$ valence bands, 
by assuming a $k$-dependent chemical potential between the $N$-th and $N + 1$-th bands at each $k$-point of the BZ.
In Fig. 2c, one can clearly find two monopoles, W2 as a source and W1 as a sink of the Berry curvature  in the positive part of BZ.
It directly confirms that W1 and W2 are a pair of WPs with opposite chirality. 
By tracing the monopole centers of Berry curvatures, we found that the coordinates of the two WPs are
W1-$(0.1024\frac{2\pi}{a},0.0128\frac{2\pi}{b},0)$ and W2-$(0.1001\frac{2\pi}{a},0.0530\frac{2\pi}{b},0)$, respectively, 
which is consistent with the band structure of Fig. 2e. 
Considering the existence of mirror symmetry in the $xz$ (glide plane) and $yz$ planes, four pairs of WPs in total can be found in the BZ (Fig. 1c). 

The existence of TRS allows us to define a  $Z_2$ topological invariant~\cite{Kane2005} on a surface with a direct energy gap.
We choose a contour $C$ that connects $S$, $\Gamma$ and $-S$ points by crossing pairs of WPs and respects the TRS (see Fig. 1b). 
Since only WPs in the $k_z=0$ plane are gapless, the $C$-$k_z$ surface is gapped everywhere. 
Thus, we can compute the $Z_2$ invariant by tracing the Wannier charge centers projected to the $z$ axis ($<Z>$) using the non-Abelian Berry connection ~\cite{Soluyanov2011, Yu2011}. 
The time-reversal partners exhibit a clear switch between the $\Gamma$ and $S$ points along the contour $C$ during time-reversal pumping~\cite{Fu2006}. 
The nontrivial $Z_2$ index is characterized by the odd number of times crossing of Wannier centers through the horizontal reference line in Fig. 2d.
It protects the existence of helical edge states along the $C$ projection to the surface BZ, which form part of the Fermi arcs between WPs with opposite chirality, if the $k_z$ direction is cut to an open surface. 

We note that the distance between W1 and W2 WPs is 4.2\% of the reciprocal lattice, which is almost six times larger than that in
WTe$_2$~\cite{Soluyanov:2015WSM2}. 
Therefore, the Fermi arcs connecting W1 and W2 should be much easier to measure by ARPES in MoTe$_2$ than in WTe$_2$.
Compared to that of WTe$_2$, the valence and conduction bands locate even closer to each other along the $\Gamma-X$ line, 
which may be attributed to the structural differences between WTe$_2$ and MoTe$_2$. 
As a consequence, pairs of WPs are separated more in MoTe$_2$ than in WTe$_2$.
Given that W1 is 6 meV above $E_F$, only a small amount of doping is required to shift the chemical potential to the WPs. 

We show the 3D energy dispersion of Weyl cones in Fig. 3.
Linear dispersions exist along all momentum directions through the nodes of W1 ad W2 points.
Because of the weak van der Waals  interactions in $z$ direction,
the Fermi velocity is much smaller in $k_z$ direction than those in $k_x-k_y$ plane.
Figures 3d-3g show the evolution of FSs in the $k_z$ = 0 plane. 
At $E_F$, the electron (orange curves) and hole (blue curves) pockets are separated by a gap.
On the energy increasing up, the electron and hole pockets extends and and shrinks, respectively. 
At $E_F$ + 6 meV, electron and hole pockets touch each other at the W1 point, as shown in Fig. 3e. 
When further increasing the energy, a new gap opens at the W1 touching point until 
the second electron hole touching appeasers at the W2 point at $E_F$ + 59 meV in Fig. 3f. 
We can clearly see that W1 and W2 WPs in MoTe$_2$ is formed by the touching points
between electron and hole pockets in the FS, which is exactly the type-II WSM proposed in Ref.~\onlinecite{Soluyanov:2015WSM2}. 
Therefore, one can also expect the appearance of Fermi arcs when the $E_F$ crosses WPs.

Next, we examine the Fermi arcs on the (001) surface. 
We consider a half-infinite surface using the iterative Green's function method~\cite{Sancho1984, Sancho1985}. 
The $k$-dependent local density of states (LDOS) are projected from the half-infinite bulk to the outermost surface unit-cell
to demonstrate the surface band structure. 
Except a small electron pocket near the ${\overline Y}$ point, the FSs mainly distribute in the center region of the surface BZ.
So we will focus on this region that includes all surface projections of WPs. 

One can see large electron and hole pockets in the FS of Fig. 4a. The hole pockets appear near  $\overline{\Gamma}$ point
and the electron pockets exist in the middle of  $\overline{\Gamma}$-${\overline X}$, which is consistent with the bulk band structure.
As presented in Figs. 4b and 4c $E_F$ + 6 meV, a Fermi arc starts from the W1 point and end at the position near the W2 point. 
Meanwhile a trivial FS co-exists, which is a closed circle with part merging into the bulk electron pocket. 
We illustrate these FSs in Fig. 4d for a simple understanding of surface states.
Because the W1 and W2 WPs do not lie at the same energy, the end point of the Fermi arc cannot exactly be the surface projection of W2. 
 As shown in Fig. 4c, the starting point is the W1 projection, while the other end point is $(0.1215\frac{2\pi}{a},0.0805\frac{2\pi}{b})$, relative far away from the W2 projection, just as expected.
As a consequence, the length of the surface Fermi arc at $E_F$ + 6 meV is much larger than the separation of W1 and W2 points in the bulk, 
$\sim 7\%$ of the reciprocal lattice, which provides an advantage for the ARPES detection.
When shifting $E_F$ to the W2 position, electron pockets expands and the hole pockets shrinks, 
and the Fermi arcs remain in the surface BZ. At $E_F$ + 59 meV, the end point of the Fermi arc becomes exactly the projection of W2 point, 
while the starting point shifts to $(0.0931\frac{2\pi}{a},0.0040\frac{2\pi}{b})$ that is close to the W1 projection. Here the Fermi arc is as long as 
$\sim 5 \%$ of the reciprocal lattice. We note that the trivial Fermi circle now merge into the W2 point, as illustrated in Fig. 4h. 
But this does not change the topology of the FS. 

We can understand further the Fermi arc states from the band structure with energy resolution. 
In Fig. 4i we can see several surface states disperse at the boundary of conduction bands and the valence bands 
along $\overline{\Gamma}$-$\overline X$. But their FSs are just closed circles, e.g. those in Figs. 4d and 4h.
Since the line connecting W1 - W2 is almost parallel to the $k_y$ axis, we plot the band structure along the $k_y$ line $\Sigma - \Sigma'$ 
(indicated in Fig. 4e).  
As shown in Fig. 4j, a surface band connects to W1 and W2 in the range of $\sim 5$ to $\sim 55$ meV above Fermi level, 
which is consistent with the bulk band structure. 
This state forms the Fermi arc observed in the FS.

We should point out that the band structure of MoTe$_2$ is very sensitive to the structural distortion. 
For example, an little strain of 1\% can annihilate two pairs of W1 WPs and leave only W2 WPs. Moreover,
The W2 WPs turns from type-II to normal type-I as a touching point by the valence band top and the conduction band bottom, 
as illustrated in Fig. 5c. 
Considering that GGA often underestimates the correlation effect and overestimates the band inversion, 
we further performed band structure calculations using the hybrid-functional method (HSE06)~\cite{HSE06}.
From the hybrid-functional calculations, we find that the original W1 WPs disappear and W2 WPs change into the type-I.
New WPs lie 45 meV above Fermi level, as the band structure showing in Fig. 5a.      
In the surface band structures (Fig. 5b), the topological surface band connects a pair of WPs with opposite chirality, 
which relates to the Fermi arc. As presented in Figs. 5d and 5e, at the energy $E=E_F+45$ meV there is
one Fermi curve starting from the WP in the positive zone of the 2D BZ and submerging into bulk state around the point of
$(0.155\frac{2\pi}{a},0.125\frac{2\pi}{b})$. This Fermi curve extends out on the left side of the bulk state and finally ends at the 
other WP with opposite chirality. This curve is the Fermi arc that connects a pair of WPs with opposite chirality.
Figure 5c illustrates the transition from type-II to type-I WSMs due to the strain or the correlation effect. 
Such phase transition should be considered when interpreting future experiments, 
due to the uncertainty of the realistic correlation effect and the lack of knowledge in the lattice contraction at low temperature.

\section{Summary}
In summary, we find the WSM state and reveal the topological Fermi arcs in the orthorhombic MoTe$_2$.
By $ab~initio$ calculations we observe four pairs of WPs with opposite chirality lying on the $k_z$ = 0 plane of the BZ.
One type of WPs are just 6 meV above the $E_F$, and the others are at $E_F$ + 59 meV, 
which can be accessed by slight electron doping.
The spacing between WP pairs is as long as $4.2 \%$ of the reciprocal lattice, six times larger than that of WTe$_2$. 
Connecting the surface projections of WPs, topological Fermi arcs exists, which calls for the experimental verification such as by ARPES.
The correlation effect or strain is expected to induce transitions from type-II to type-I WSMs in MoTe$_2$, 
which are calling for experimental verifications.

\begin{acknowledgments}
We are grateful for Y. P. Qi and B. A. Bernevig for helpful discussions. This work  was financially supported  by the  Deutsche Forschungsgemeinschaft DFG  (Project No. EB 518/1-1 of DFG-SPP 1666 "Topological Insulators", and SFB 1143) 
and by the ERC (Advanced Grant No. 291472 "Idea Heusler").  
\end{acknowledgments}

%

\end{document}